# MultiAmdahl: Optimal Resource Allocation in Heterogeneous Architectures

L. Yavits, A. Morad, U. Weiser, R. Ginosar

**Abstract**— Future multiprocessor chips will integrate many different units, each tailored to a specific computation. When designing such a system, the chip architect must decide how to distribute limited system resources such as area, power, and energy among the computational units. We extend MultiAmdahl, an analytical optimization technique for resource allocation in heterogeneous architectures, for energy optimality under a variety of constant system power scenarios. We conclude that reduction in constant system power should be met by reallocating resources from general-purpose computing to heterogeneous accelerator-dominated computing, to keep the overall energy consumption at a minimum. We extend this conclusion to offer an intuition regarding energy-optimal resource allocation in data center computing.

**Index Terms**— Heterogeneous computing, high performance computing, optimal resource allocation, energy optimization, data center computing.

---

## 1 INTRODUCTION

Heterogeneous multiprocessor chips integrate a large number of different computational units: these include full-blown latency oriented cores for sequential processing, massively parallel SIMD accelerators such as vector (VPU) or graphic processing (GPU) units, media accelerators, and application-specific integrated circuits (ASICs). The last are designed and optimized for a particular workload, and therefore are much more efficient than general-purpose units [4].

Heterogeneity also has a price; chips have limited physical resources, such as die area or average/peak power/energy. All units on the chip compete for the shared resources, and the share that each unit receives is limited. When a single chip contains multiple different units with different roles, it is up to the architect to distribute the resources among the different units. To reach an optimal solution, the architect should take into account the efficiency of these units as well as the workload.

Fig 1 schematically illustrates the application range and power efficiency (performance to power ratio) over this range for each computational unit of a heterogeneous chip. For instance, the general-purpose CPU is designed to implement a wide range of applications, with relatively low power efficiency, whereas designated accelerators such as a Fast Fourier Transform (FFT), Black-Scholes (BSC), or Dense Matrix Multiplication (DMM) provide high power efficiency for a narrow range of applications.

In this paper, we extend the MultiAmdahl [12] framework for energy optimality under constant system power and study its effect on optimal resource allocation in heterogeneous architectures. We analyze a high level High Performance Computing (HPC) architecture and present a closed-form solution for optimal resource allocation. We then extend this line of reasoning to intuit how resources might be allocated in in data centers in an energy-optimal manner. Finally, we provide insight into optimal resource distribution in a large scale heterogeneous architecture featuring a number of special purpose accelerators, under various levels of constant system power.

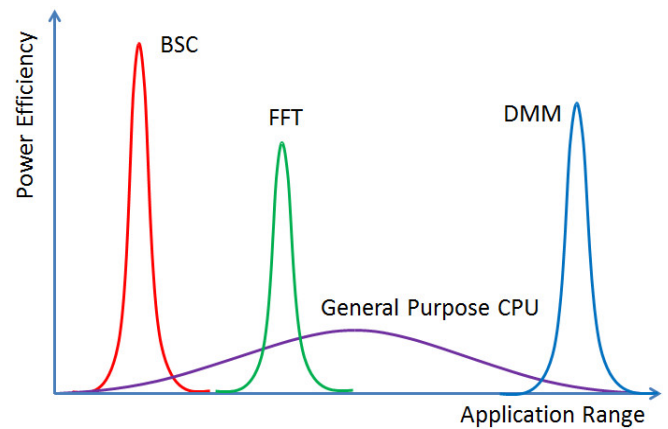

Fig 1. Power efficiency *vs.* application range

The rest of this paper is organized as follows. Section 2 discusses the resource allocation problem. Section 3 introduces the optimization framework. Section 40 presents the results of HPC architecture optimization. Section 5 discusses optimization of a heterogeneous architecture with many accelerators, and section 6 presents our conclusions.

## 2 THE RESOURCE ALLOCATION PROBLEM

Consider a heterogeneous system composed of several different units, each with a different role. A common resource (for example, the die area) is shared among the different units. The system architect needs to distribute the resource among the available units. Increasing the size of one unit improves its performance, but also force all other units to share a smaller area, thus reducing their performance. To find the optimal resource allocation, four parameters must be considered.

---
* *Leonid Yavits, E-mail: yavits@technion.ac.il.*
* *Amir Morad, E-mail: amirm@tx.technion.ac.il.*
* *Uri Weiser, E-mail: uri.weiser@ee.technion.ac.il.*
* *Ran Ginosar, E-mail: ran@ee.technion.ac.il.*
* *Authors are with the Department of Electrical Engineering, Technion-Israel Institute of Technology, Haifa 3200000, Israel.*

The first parameter is the overall resource constraint. In this paper, we focus on the die area constraint. We can also introduce additional resource constraints, such as peak power or energy.

The second parameter that must be considered is the accelerator function, meaning how the additional resource (e.g., area, power, or energy) assigned to a given accelerator is translated into a better target performance. For example, doubling the size of a massively parallel vector accelerator (VPU) is likely to nearly double its potential performance. A latency-oriented sequential CPU, however, is not likely to exhibit the same improvement when its size is doubled.

The system architect must also decide how to allocate the resource within each unit: how to divide the unit's area between the processing unit and the memory, for example [1]. Accelerator functions hide these details, always providing the optimal output of the unit given the amount of resource it was assigned.

A third parameter of the optimization problem is the workload of the system, and especially its distribution among the different units. The workload determines how much work is assigned to each of the accelerators. In common scenarios, the accelerator-functions are known a priori, while specific details of the workload are only estimated.

The last parameter is the design goal the system architect seeks to optimize for.

A design constraint must be met, while the design goal presents one design as preferable to another. A design may have several constraints (power, area, memory bandwidth [7]), but only a single goal. For example, a real-time system may require optimization of its performance. On the opposite side, as energy-related costs rise, a high-performance computing (HPC) system designed for data centers may require optimizing its energy consumption.

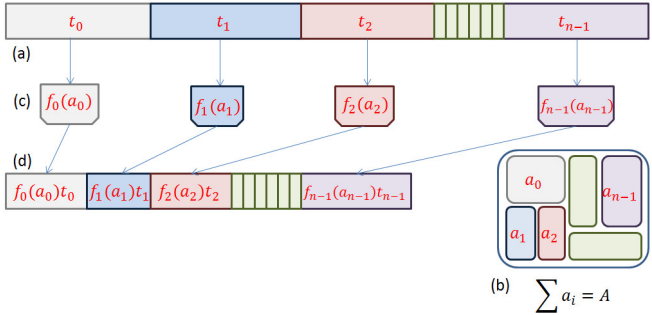

Fig 2. MultiAmdahl framework: (a) Code segment execution on a reference CPU; (b) Die area division; (c) Accelerator performance; (d) Execution time using the accelerators

## 3 OPTIMIZATION FRAMEWORK

In this section, we formulate the optimization problem on the basis of the four factors described in the previous section.

Workload — We divide the workload into $n$ different execution segments. Each segment represents the aggregated amount of work to be executed by a specific accelerator of a heterogeneous chip (Fig 2 (a)). We define $t_i$ as the execution time of segment $i$ on a reference CPU. Thus, the total execution time of the workload on the reference CPU is $\sum_{i=0}^{n-1} t_i$.

Area Constraint — the chip area is divided among $n$ hardware computational units (see Fig 2(b)), where each unit $i$ executes segment $i$. We denote by $a_i$ the chip area that is allocated to unit $i$. The sum of the areas assigned to all units is bounded by the total chip area $A$:

$$\sum_{i=0}^{n-1} a_i = A \quad (1)$$

Accelerator functions — Accelerator function $f_i$ (see Fig 2(c)) represents the inverted speedup of the $i_{th}$ accelerator as a function of the area $a_i$ assigned to the accelerator. Therefore, using the $i^{th}$ accelerator, the execution time of segment $i$ is $t_i \cdot f_i(a_i)$ [12]. When relevant, this function should take into account the migration overhead of the appropriate accelerator [2]. To simplify the analysis, we assume that the accelerator functions are strictly decreasing, convex, and continuously differentiable. A simple form of analytical function that satisfies this requirement is a power-law function [12]:

$$f_i(a_i) = a_i^\alpha \quad (2)$$

where $a_i$ is the resource assigned to the processor, and $\alpha$ is a negative constant coefficient.

Pollack [5] suggests that the performance of a single-core CPU can be modeled as the square root of the assigned area. In our notations, $\alpha = -0.5$ for a full-blown CPU core. On the other end of the core complexity scale is a massively parallel vector accelerator (VPU), comprising a multitude of baseline throughput oriented cores. We model the performance of such a massively parallel accelerator as (slightly) sublinear to the number of cores (or resource amount). Sublinear scaling reflects the damping effect of inter-core communication [8]. In our research, we span the value of $\alpha$ for massively parallel accelerators from $-1$ ("embarrassing" parallelism, where performance scales linearly to the number of cores) to $-0.75$, and discover that varying $\alpha$ does not significantly affect the outcome of our analysis.

### 3.1 Optimizing Delay

Our first design goal is to minimize the total execution time when using accelerators (see Fig 2(d)). Adding the area constraint, we arrive at the following optimization problem:

$$\begin{cases} \text{Minimize:} & \sum_{i=0}^{n-1} f_i(a_i) t_i \\ \text{Subject to:} & \sum_{i=0}^{n-1} a_i = A \end{cases} \quad (3)$$

This is a convex optimization problem, which can be solved using Lagrange multipliers [11]:

$$\text{Minimize: } \Lambda(\bar{a}, \lambda) = \sum_{i=0}^{n-1} f_i(a_i)t_i - \lambda(\sum_{i=0}^{n-1} a_i - A) \quad (4)$$

An optimal solution satisfies:

$$\begin{cases} \frac{\partial \Lambda}{\partial a_i} = 0, \forall\, 0 \leq i \leq n-1 \\ \frac{\partial \Lambda}{\partial \lambda} = 0 \end{cases} \quad (5)$$

Solving for arbitrary index $i$:

$$f'_i(a_i)t_i - \lambda = 0 \quad (6)$$

Solving for arbitrary index $j$, we obtain the dual formula:

$$f'_j(a_j)t_j - \lambda = 0, \quad (7)$$

and so the optimal solution is obtained when:

$$f'_i(a_i)t_i = f'_j(a_j)t_j, \quad (8)$$

where $f'_i(a_i)$ is the derivative of the $i_{th}$ accelerator function, $a_i$ is the area assigned to this accelerator, and $t_i$ is the execution time of this segment on the reference CPU.

The intuition behind this rule is that in an optimal solution, any infinitesimal addition to the area creates the same improvement in the total execution time, regardless of the accelerator it is assigned to.

### 3.2 Optimizing Energy

Our second design goal is to minimize the total energy consumption when using accelerators. Energy consumption of a heterogeneous system can be expressed as follows:

$$E = \left(\sum_{i=0}^{n-1} p_i(a_i) f_i(a_i) t_i\right) + E_{sys}$$
$$= \sum_{i=0}^{n-1} [p_i(a_i) + P_{sys}] f_i(a_i) t_i, \quad (9)$$

where $p_i(a_i)$ is the dynamic power consumption of the $i^{th}$ accelerator, and $E_{sys}$ and $P_{sys}$ are constant system energy and power respectively. $P_{sys}$ may include, for example, static (leakage) power and DRAM refresh power. On a larger scale, $P_{sys}$ can also account for cooling, UPS, lighting and other infrastructure related power components of data centers.

Following Chung [4], we model the dynamic power consumption of a unit as a power law of its area:

$$p_i(a_i) = a_i^\beta \quad (10)$$

where $a_i$ is the resource assigned to the unit, and $\beta$ is a positive constant coefficient. Grochowsky and Annavaram [3] suggest that for a single-core CPU, $\beta = 0.875$. For a massively parallel accelerator, however, we scale power consumption superlinearly to the number of cores (or resource amount). Such scaling reflects the excess power spent on inter-core communication, which itself scales superlinearly to the number of cores [8]. In our research, we span the value of $\beta$ for massively parallel accelerators from 1 to 1.25 and discover that varying $\beta$ does not significantly change the outcome of our analysis.

Adding the area constraint, we can write the energy optimization problem as follows:

Minimize:

$$\Lambda = \sum_{i=0}^{n-1} [p_i(a_i) + P_{sys}] f_i(a_i) t_i - \lambda(\sum_{i=0}^{n-1} a_i - A) \quad (11)$$

Similarly to (8), the optimal solution is obtained when:

$$\{[p_i(a_i) + P_{sys}]f_i(a_i)\}'t_i = \{[p_j(a_j) + P_{sys}]f_j(a_j)\}'t_j \quad (12)$$

## 4 OPTIMIZING A HETEROGENEOUS HPC ARCHITECTURE

Consider a typical HPC architecture consisting of a number of general purpose CPU cores, a massively parallel vector accelerator, and memory. This top level organization is shared by Intel Core™, NVidia Echelon, and other contemporary HPC architectures. In this section, we apply the MultiAmdahl framework to optimize the allocation of the chip area resource between the general purpose CPU cores and the massively parallel vector accelerator (VPU).

Equations (12) and (1) can be rewritten to reflect the area allocation between $a_{CPU}$ and $a_{VPU}$:

$$\begin{cases} [\gamma_{CPU} a_{CPU}^{\gamma_{CPU}-1} + P_{sys}\alpha_{CPU} a_{CPU}^{\alpha_{CPU}-1}] t_{CPU} = \\ [\gamma_{VPU} a_{VPU}^{\gamma_{VPU}-1} + P_{sys}\alpha_{VPU} a_{VPU}^{\alpha_{VPU}-1}] t_{VPU}, \\ a_{CPU} + a_{VPU} = A \end{cases} \quad (13)$$

where $\gamma_{CPU} = \alpha_{CPU} + \beta_{CPU}$, $\gamma_{VPU} = \alpha_{VPU} + \beta_{VPU}$, $\alpha_{CPU} = -0.5$, $\beta_{CPU} = 0.875$, $-1 \geq \alpha_{VPU} \geq -0.75$ and $1 \leq \beta_{VPU} \leq 1.25$.

The intuition behind (13) is that in an optimal solution, any infinitesimal addition to the area creates the same improvement in the total energy consumption (dynamic plus system), regardless of the accelerator it is assigned to.

Solving (13) for $a_{CPU}$ and $a_{VPU}$ yields the energy-optimal partitioning of the chip area resource to the CPU and VPU. Fig 3 shows the normalized energy of an HPC as a function of $a_{CPU}$ relative to the total chip area budget $A$ for different levels of constant system power. We assume $P_{sys}$=2%, 10%, 40% and 95% of the total power budget (which is derived from the chip area budget $A$ (10)), and $t_{CPU} = t_{VPU}$ (Amdahl's parallelization factor of 0.5). For each $P_{sys}$ level, we show the optimal CPU area allocation (connected by the red curve).

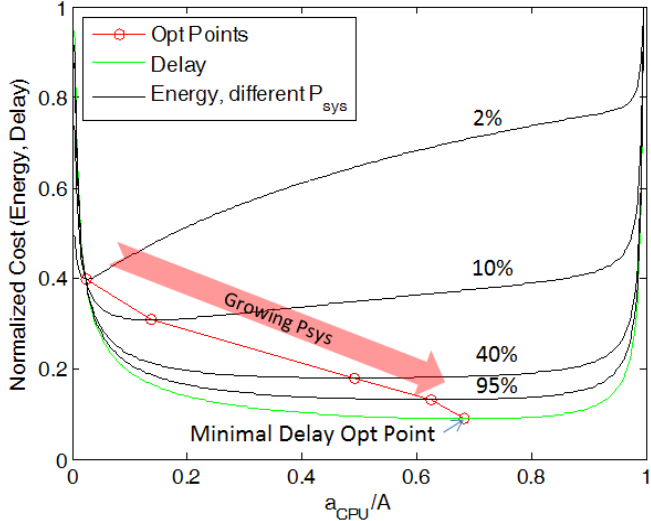

Fig 3. Normalized objective (cost) function *vs.* area allocation for CPU for different $P_{sys}$

When $P_{sys}$ is very low (2% of the system power budget), the energy-optimal area allocation to the CPU is also very low. The intuition behind this is as follows: to conserve the overall energy, the least power efficient unit (the unit with the highest $p_i(a_i)f_i(a_i)$, i.e., the CPU) should run as slowly as possible. This is achieved by allocating the smallest possible area budget to it. As $P_{sys}$ increases, the optimal area allocation to the CPU grows as well. This happens because with growing $P_{sys}$, the relative weight of the second summand of (9) also increases. To reduce this component of the total energy consumption, the execution needs to be completed as soon as possible; hence, the slowest unit (CPU) needs to be sped up, which is achieved by reducing $f_{CPU}(a_{CPU})$, by increasing $a_{CPU}$.

Eventually, if almost the entire energy budget is spent on system infrastructure as opposed to computation, the optimal CPU allocation approaches the optimal point of minimal delay (Section 3.1), which is also shown in Fig 3 for reference.

The relationship between the optimal area allocation in a heterogeneous HPC architecture and constant system power is summarized in Fig 4. The energy-optimal CPU area allocation increases (and VPU area decreases) as system power grows, approaching the delay-optimal area allocation. Generally, the following holds:

$$\lim_{P_{sys}\to\infty} a_{CPU\ min\ energy} = a_{CPU\ min\ delay} \qquad (14)$$

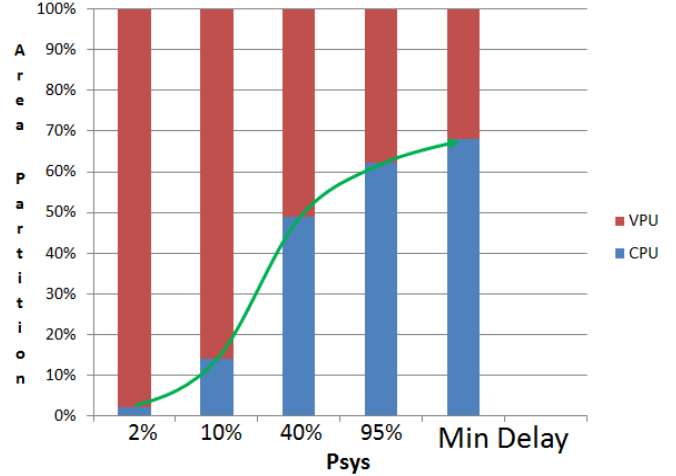

Fig 4. Area allocation in HPC architecture *vs.* constant system power

Next we consider the impact of this result on area optimization in future heterogeneous HPC architectures and provide an intuition as to how to allocate resources in data center computing.

### 4.1 Effect of System Power on Resource Allocation in HPC Architecture

In addition to its computational part (CPU and VPU), an HPC architecture features large shared memory. Today, such memory is typically DRAM, increasingly integrated with the computing die in the same package (for example, using 2.5D or 3D integration), as depicted in Fig 5(a). 3D DRAM may exhibit fairly significant static power (leakage and refresh), potentially biasing the optimal area allocation between the CPU and VPU in favor of the CPU (*cf.* Fig 4).

However, as CMOS feature scaling slows down, conventional memory technology such as DRAM experiences scalability problems. In response, resistive memory (ReRAM) technologies are being explored. ReRAM stores information by modulating the resistance of nanoscale storage elements. One of ReRAM's noticeable advantages is non-volatility, which obviates the need for refresh and provides near-zero leakage power.

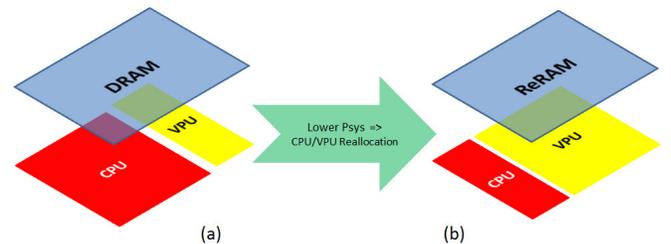

Fig 5. Effect of constant system power on resource allocation in HPC architecture

If ReRAM replaces DRAM in the HPC architecture in question, as depicted in Fig 5(b), the system power is likely to decrease quite considerably. To keep the resulting design at the point of optimal energy, a reallocation of the area resource is required. Such a new allocation is likely to favor VPU at the expense of CPU. We believe this insight could be quite important for computer architects.

## 4.2 Heterogeneity in Data Centers

This line of reasoning can be expanded from an HPC module to the level of a data center. The power consumption of a typical data center consists of IT power (spent on computation and data retention) and infrastructure power (cooling, UPS, lighting, power distribution etc.). The infrastructure power can be quite significant: cooling power alone may reach 45% of the entire power budget of a data center.

Similarly to IT power, which consists of dynamic and static (leakage) components, the infrastructure power can also be divided into two components. The first is constant power (such as lighting). More significant, however, is the variable component of the infrastructure power, which depends on IT power (for example, cooling power, which is almost linearly dependent on IT power). To reflect this division within the total data center energy, we rewrite (9) as follows:

$$E = \sum_{i=0}^{n-1}[p_i(a_i) + P_{Infrastructure}]f_i(a_i)t_i$$
$$= \sum_{i=0}^{n-1}[wp_i(a_i) + P_{const}]f_i(a_i)t_i \quad (15)$$

where $P_{Infrastructure} = P_{const} + O(p_i(a_i))$, where $P_{const}$ is the constant component of the infrastructure power, while the IT power-dependent component $O(p_i(a_i))$ is absorbed in $wp_i(a_i)$, where $w \geq 1$ is constant.

Applying our energy optimization framework to (15) leads to the following conclusion. If the constant component of data center energy is significant, the energy-optimal data center should favor general-purpose computing (full-blown CPU cores). However, if the constant power is reduced, the energy-optimal point shifts towards the heterogeneous approach.

In general, the challenge of minimizing the constant system power should be met by reallocating resources from general-purpose CPU based computing to heterogeneous accelerator based computing, to keep the overall energy consumption of the data center at a minimum.

## 5 OPTIMIZING A MULTI-ACCELERATOR HETEROGENEOUS ARCHITECTURE

While in the previous section we considered optimal resource allocation between a general purpose CPU and a massively parallel vector accelerator, in this section we focus on a heterogeneous architecture featuring a general purpose CPU and a multitude of ASIC accelerators. We follow Chung *et al.* [4] and consider an architecture composed of a CPU and Dense Matrix Multiplication, FFT16, FFT 1024, and Black-Scholes option pricing accelerators. The workload is composed of equal-runtime dense matrix multiplication, FFT16, FFT 1024 and Black-Scholes routines [4], where 10% of the runtime of each routine is assumed to be executed on the CPU and the rest on their dedicated accelerators respectively. As earlier, only one unit in the chip is active at a time. The accelerator functions take the form:

$$f_i(a_i) = \frac{a_i^\alpha}{e_i} \quad (16)$$

where the area efficiency factor $e_i$ is as provided by Chung *et al.* [4], and summarized in TABLE 1.

TABLE 1
ACCELERATOR EFFICIENCIES [4]

| Benchmark | Area Efficiency Factor $e_i$ |
|---|---|
| Dense Matrix Multiplication | 39 |
| FFT1024 | 692 |
| FFT16 | 2804 |
| BlackScholes | 24 |
| CPU | 1 |

Fig 6 presents the energy-optimal area allocation in such a heterogeneous chip for different values of constant system power. When $P_{sys}$ is very low, the allocation to the most area-inefficient unit (the CPU) is minimal. The rest of the area budget is assigned to the accelerators in accordance with their area efficiency: the largest portion is allocated to the most area-efficient accelerator, and so on.

However, as system power increases, the energy-optimal allocation of area resources changes significantly. The allocation to area-inefficient units grows very quickly (with most area being eventually allocated to the CPU), while the allocation to the most area-efficient accelerator drops at the same rate.

The rightmost column of the Fig 6 shows the delay-optimal area allocation (Section 3.1 as well as [12]). Similarly to the HPC architecture case, the energy-optimal area allocation in the heterogeneous multi-accelerator architecture approximates the delay-optimal allocation as system power increases.

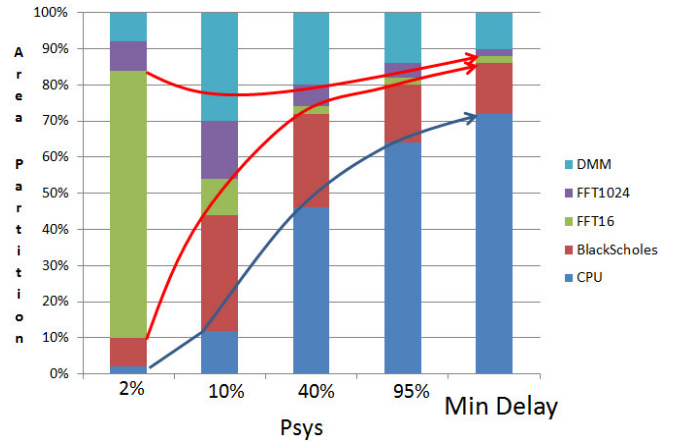

Fig 6. Energy-optimal allocation vs. $P_{sys}$; the rightmost column shows the delay-optimal allocation [12].

## 6 CONCLUSIONS

We have extended the MultiAmdahl analytical delay optimization framework to energy-optimal resource allocation in heterogeneous architectures of processors and data centers. Our technique relies on modeling the performance and power consumption of each unit as a function of the resources it uses, and setting the limited resource as an optimization constraint. We then find the optimal solution using Lagrange multipliers. The following contributions with regard to the MultiAmdahl framework are made:

1. We illustrate how the optimal performance or energy

consumption under a resource constraint is achieved when the absolute marginal outcome is equal for all computing modules.

2. We introduce a closed form analytical solution for optimal resource allocation in heterogeneous architectures.

3. We observe that as the total system power increases, the optimal energy allocation approaches the allocation for minimum delay.

4. We conclude that the challenge of minimizing the constant system power should be met by reallocating resources from general-purpose CPU based computing to heterogeneous accelerator based computing, to keep the overall energy consumption at a minimum.

5. We show how the change in constant system power, for example, the transitioning from DRAM to Resistive RAM, may affect the resource allocation in High Performance Computing architectures.

6. We apply similar reasoning to intuit how resources might be allocated in data centers in an energy-optimal manner: as the power efficiency of data centers improves, they should shift away from general purpose computing to heterogeneous accelerator-dominated computing, to maintain energy optimality.

In summary, the MultiAmdahl framework provides a wide range of optimization opportunities, and a multi-functional tool for an early stage analytic exploration of heterogeneous computing design space.

## ACKNOWLEDGMENT

We thank Ronny Ronen for his valuable input. This research was partially supported by the Intel Collaborative Research Institute for Computational Intelligence (ICRI-CI) and Hasso-Plattner-Institut.

## REFERENCES

[1] Morad, A., Yavits, L. and Ginosar, R., 2014, September. Convex optimization of resource allocation in asymmetric and heterogeneous SoC. InPower and Timing Modeling, Optimization and Simulation (PATMOS), 2014 24th International Workshop on (pp. 1-8). IEEE..

[2] Morad, A., Morad, T.Y., Leonid, Y., Ginosar, R. and Weiser, U., 2014. Generalized MultiAmdahl: optimization of heterogeneous multi-accelerator SoC. Computer Architecture Letters, 13(1), pp.37-40.

[3] E. Grochowski and M. Annavaram, "Energy per instruction trends in Intel microprocessors," Technology@Intel Magazine 4(3), 2006.

[4] Chung, E.S., Milder, P.A., Hoe, J.C. and Mai, K., 2010, December. Single-chip heterogeneous computing: Does the future include custom logic, FPGAs, and GPGPUs?. In Proceedings of the 2010 43rd Annual IEEE/ACM International Symposium on Microarchitecture (pp. 225-236). IEEE Computer Society.

[5] F. J. Pollack, "New microarchitecture challenges in the coming generations of CMOS process technologies" (keynote address), Proc. MICRO-32, 1999.

[6] G. M. Amdahl, "Validity of the single processor approach to achieving large scale computing capabilities," Proc. AFIPS-67, pp. 483-485.

[7] Yavits, L., Morad, A. and Ginosar, R., 2014", Cache hierarchy optimization", IEEE Computer Architecture Letters, 13(2), pp.69-72.

[8] Yavits, L., Morad, A. and Ginosar, R., 2014. The effect of communication and synchronization on Amdahl's law in multicore systems. Parallel Computing, 40(1), pp.1-16.

[9] Yavits. L, A. Morad, R. Ginosar, "Computer Architecture with Associative Processor Replacing Last Level Cache and SIMD Accelerator", IEEE Transactions on Computers, 2015, vol. 64, issue 2, pp 368 - 381

[10] Kumar, R., Farkas, K.I., Jouppi, N.P., Ranganathan, P. and Tullsen, D.M., 2003, December. Single-ISA heterogeneous multi-core architectures: The potential for processor power reduction. In Microarchitecture, 2003. MICRO-36. Proceedings. 36th Annual IEEE/ACM International Symposium on (pp. 81-92). IEEE.

[11] R. Rockafellar, "Lagrange multipliers and optimality," SIAM Review, vol. 35, 1993, pp. 183-288.

[12] Zidenberg, T., Keslassy, I. and Weiser, U., 2012. MultiAmdahl: How should I divide my heterogenous chip?. Computer Architecture Letters, 11(2), pp.65-68.